\documentclass[twocolumn]{aastex62}

\usepackage{verbatim}
\usepackage{enumitem}

\usepackage{nicefrac}
\graphicspath{{./}{figures/}}

\received{January 1, 2018}
\revised{January 7, 2018}
\accepted{\today}
\submitjournal{ApJ}

\shorttitle{The isotropic center of NGC\,5419}
\shortauthors{Neureiter et al.}

\begin{document}

\title{The isotropic center of NGC\,5419 - A core in formation?}

\correspondingauthor{Bianca Neureiter}
\email{bneu@mpe.mpg.de}

\author[0000-0001-6564-9693]{Bianca Neureiter}
\affil{Max-Planck-Institut für extraterrestrische Physik,Giessenbachstrasse 1, D-85748 Garching, Germany}
\affil{Universitäts-Sternwarte München, Scheinerstrasse 1, D-81679 München, Germany}

\author{Jens Thomas}
\affil{Max-Planck-Institut für extraterrestrische Physik,Giessenbachstrasse 1, D-85748 Garching, Germany}
\affil{Universitäts-Sternwarte München, Scheinerstrasse 1, D-81679 München, Germany}

\author{Antti Rantala}
\affil{Max-Planck-Institut f\"ur Astrophysik, Karl-Schwarzschild-Str. 1, D-85748, Garching, Germany }

\author{Thorsten Naab}
\affil{Max-Planck-Institut f\"ur Astrophysik, Karl-Schwarzschild-Str. 1, D-85748, Garching, Germany }

\author{Kianusch Mehrgan}
\affil{Max-Planck-Institut für extraterrestrische Physik,Giessenbachstrasse 1, D-85748 Garching, Germany}
\affil{Universitäts-Sternwarte München, Scheinerstrasse 1, D-81679 München, Germany}

\author{Roberto Saglia}
\affil{Max-Planck-Institut für extraterrestrische Physik,Giessenbachstrasse 1, D-85748 Garching, Germany}
\affil{Universitäts-Sternwarte München, Scheinerstrasse 1, D-81679 München, Germany}

\author{Stefano de Nicola}
\affil{Max-Planck-Institut für extraterrestrische Physik,Giessenbachstrasse 1, D-85748 Garching, Germany}
\affil{Universitäts-Sternwarte München, Scheinerstrasse 1, D-81679 München, Germany}

\author{Ralf Bender}
\affil{Max-Planck-Institut für extraterrestrische Physik,Giessenbachstrasse 1, D-85748 Garching, Germany}
\affil{Universitäts-Sternwarte München, Scheinerstrasse 1, D-81679 München, Germany}

\begin{abstract}
With its cored surface brightness profile, the elliptical galaxy NGC\,5419 appears as a typical high-mass early-type galaxy (ETG). However, the galaxy hosts two distinct nuclei in its center.
We use high-signal MUSE (Multi-Unit Spectroscopic Explorer\footnote{Based on observations collected at the European Organisation \\ for Astronomical Research in the Southern Hemisphere under ESO \\program 099.B-0193(A).}) spectral observations and novel triaxial dynamical orbit models to reveal a surprisingly isotropic central orbit distribution in NGC\,5419.
Recent collisionless simulations of merging massive ETGs suggest a two-phase core formation model, in which the low-density stellar core forms rapidly by supermassive black holes (SMBHs) sinking into the center due to dynamical friction. Only afterwards the SMBHs form a hard binary and the black hole scouring process slowly changes the central orbit distribution from isotropic to tangential. 
The observed cored density profile, the double nucleus and the isotropic center of NGC\,5419 together thus point to an intermediate evolutionary state where the first phase of the core formation has taken place, yet the scouring process is only beginning. This implies that the double nucleus is a SMBH binary.
Our triaxial dynamical models indicate a total mass of the two SMBHs in the center of NGC\,5419 of $M_\mathrm{BH} = (1.0 \pm 0.08) 10^{10} M_\odot$. 
Moreover, we find that NGC\,5419's complex kinematically distinct core (KDC) can be explained by a coherent flip of the orbital rotation direction of stars on tube orbits at \textcolor{black}{$\sim 3$kpc} distance from the galaxy center together with projection effects. This is also in agreement with merger simulations hosting SMBHs in the same mass regime.
\end{abstract}


\section{Introduction} \label{sec:intro}

NGC\,5419 is the dominant galaxy of the poor cluster Abell S753 observable in the Centaurus constellation.
With a brightness of $M_V=-23.1$, NGC\,5419 belongs to the class of massive early type galaxies (ETGs), which are thought to undergo gas-poor galaxy merging processes at least at late evolutionary phases (see e.g. \citealt{Bender88_b,Bender92,Kormendy96,Moster18}). 
As a result of this, ETGs at the high-mass end characteristically show a central cored surface brightness (SB) profile \citep{Nieto91a,Nieto91b,Faber97,Lauer05,Lauer07} that indicates a light-deficit \citep{Kormendy09,Kormendy13}. Such cores naturally form in gas-poor mergers with supermassive black holes (SMBHs) (e.g. \citealt{Begelman80, Hills80, Ebisuzaki91, Milosavljevic01,Merritt06, Rantala18,Nasim21}).
By using high resolution galaxy merger simulations with SMBHs \citet{Rantala18} and \citet{Frigo21} recently revealed that the core formation actually happens in two phases: First, dynamical friction causes the two SMBHs of the progenitor galaxies to sink to the centre of the merger remnant. This causes the surrounding stars to move to larger radii, happens rapidly (some tens of Myrs) and is the main driver of the formation of the shallow central stellar density core. At the time when the two SMBHs form a tightly bound (hard) binary, most of the stellar density core structure is already in place. 
Afterwards, in a second and slower phase, slingshot interactions with the formed SMBH binary kick out stars, which are predominantly on radial orbits and get close enough to the binary. The result is a characteristic tangentially-biased central orbit structure. This second, slower evolutionary phase (several 100 Myrs to up to 1 Gyr) flattens the SB profile only slightly. Inner tangential anisotropy is common in ETGs (e.g. \citealt{Gebhardt03, McConnell12}). The most massive ETGs, in particular, have a very uniform anisotropy structure near their cores that matches very well with the predictions of the SMBH binary model \citep{Thomas14,Rantala18,Mehrgan19}.

While NGC\,5419 with its central SB core appears as a typical ETG at first glance, 
this galaxy is clearly distinct from other elliptical core galaxies as Hubble Space Telescope (HST) as well as SINFONI observations reveal a double nucleus in its center (\citealt{Lauer05, Capetti05, Lena14, Mazzalay16}). Besides a low-luminosity active galactic nucleus at the center of the galaxy (\citealt{Goss87,Subrahmanyan03,Balmaverde06}), NGC\,5419 hosts a second nucleus about $\sim0.25$ arcsec away from the AGN (\citealt{Lauer05, Capetti05, Lena14, Mazzalay16}).
Assuming a distance of $56.2$ Mpc for NGC\,5419 \citep{Mazzalay16}, this corresponds to a separation of $\sim70$pc. High-resolution SINFONI spectra reveal a high stellar velocity dispersion around both nuclei and between them, suggesting that each nucleus carries a SMBH and their masses are almost equal \citep{Mazzalay16}. Axisymmetric dynamical models indicate a combined SMBH mass of $M_\mathrm{BH} = 7.2^{+2.7}_{-1.9} \times 10^9 M_\odot$ \citep{Mazzalay16}. Together with the observed radial separation of the two nuclei this implies that if the double nucleus is indeed a SMBH binary, then it is just about to become a hard binary.

With this, NGC\,5419 could be the first ETG observed just at the transition between the two core formation phases described above: The rapid formation of the shallow density core through dynamical friction appears to have already taken place. However, the observed double nucleus suggests that the slower BH scouring process causing the characteristic tangentially biased central orbit distribution, might have just begun. \\
The key to reveal the evolutionary stage of NGC\,5419 is therefore to get precise measurements of NGC\,5419's anisotropy profile. In this paper, we provide a new dynamical analysis of NGC\,5419 that extends the previous models by \citet{Mazzalay16} in several respects: we use additional 2D stellar kinematics based on MUSE observations \citep{Mehrgan19}. We use non-parametric line-of-sight velocity distributions (LOSVDs) derived with a novel non-parametric spectral fitting code called WINGFIT (Thomas et al., in prep). Our new dynamical models are triaxial rather than axisymmetric. They are based on a newly developed non-parametric deprojection routine \texttt{SHAPE3D} \citep{deNicola20} and the new triaxial Schwarzschild code \texttt{SMART} \citep{Neureiter21}. The current version of \texttt{SMART} uses a recently developed model selection framework \citep{Lipka21,Thomas22} that avoids potential biases in $\chi^2$-based models. \\
This Letter is organized as follows: In Section~\ref{sec:Observations} we briefly present the observations. In Section~\ref{sec:Triaxial Dynamical Modeling} we explain the used triaxial deprojection and dynamical modeling machinery. In Section~\ref{sec:Results} we present our findings and results, which we discuss and summarize in Sections~\ref{sec: Discussion} and~\ref{sec: Conclusion}.

\section{Observations} \label{sec:Observations}

\subsection{Surface Brightness} \label{sec:Surface Brightness}
For our current study we use the SB data and analysis as described in~\citet{Mazzalay16}, which is based on archived HST/WFPC2 F555W and $3.6 \mu m$ Spitzer IRAC1 images. After masking the two central nuclei and matching the HST and Spitzer data (for a more detailed description we refer to~\citealt{Mazzalay16}), the elliptical isophotes were determined by using the \texttt{IRAF} task \texttt{ELLIPSE} \citep{Jedrzejewski87}. The corresponding surface brightness, ellipticity and position angle profiles are discussed in more detail in Section~\ref{sec: Shape recovery}.

\subsection{Non-parametric Stellar Kinematics} \label{sec:Kinematics}
The stellar kinematical data for the dynamical modelling is based on spectroscopic observations for NGC\,5419 using MUSE at the Very Large Telescope (VLT). NGC\,5419 is part of a larger sample\footnote{ESO program 099.B-0193(A), P.I. J. Thomas} of massive ETGs for which we obtained MUSE spectra with high signal-to-noise ratio (SNR). The data analysis and stellar kinematical measurements for the whole program are described in \textcolor{black}{detail in \citet{Mehrgan23}}. 
Our kinematic analysis is identical to what is explained in that paper only that we bin for a SNR that is roughly twice as high (to reduce the number of kinematic bins by a factor of two). \textcolor{black}{The resulting spatial resolution is sufficiently high in order to produce robust and reliable results (see Sections~\ref{sec: Mass recovery} and~\ref{sec: Anisotropy recovery}) and it allows for an optimized usage of computational resources.}
To extract non-parametric LOSVDs for the dynamical models we use the non-parametric modelling code WINGFIT. For illustration, we show maps of the velocity, dispersion and Gauss-Hermite coefficients \citep{Gerhard93,vanderMarel93} derived from the non-parametric LOSVDs of NGC\,5419 in Figure~\ref{fig:velmap}. More details about the kinematics and plots with the entire LOSVDs can be found in \citet{Mehrgan23}.  \\
To test higher spatial resolution in the center, we performed an additional modeling analysis of a combined data-set of the MUSE kinematics together with central adaptive-optics (AO) assisted near-infrared observations from the SINFONI integral field spectrograph at the VLT. We use the non-parametric LOSVDs derived from the SINFONI data by \citet{Mazzalay16}. The FOV of these data covers $\sim (3\times 3)$ arcsec and the AO operation enabled a spatial resolution of $\sim0.2$ arcsec.  Similar to the study in~\citet{Mazzalay16} we exclude the innermost Voronoi bins with $r<0.35$ arcsec, where the galaxy hosts its double nucleus.\\

\section{Triaxial Dynamical Modeling} \label{sec:Triaxial Dynamical Modeling}
NGC\,5419 is the first galaxy which we analyse with our new triaxial codes \texttt{SHAPE3D} for the deprojection and \texttt{SMART} for the Schwarzschild orbit models. The two codes are described in detail in \citet{deNicola20} and \cite{Neureiter21}. The Schwarzschild models \citep{Schwarzschild79} employ a new model selection framework extension, in which a generalised information criterion $\mathrm{AIC_p} = \chi^2 + 2 m_\mathrm{eff}$ rather than the usual $\chi^2$ gets minimized \citep{Lipka21,Thomas22}. By taking the individual model's degrees of freedom $m_\mathrm{eff}$ into account, bias in the evaluation of different mass models gets avoided. In \citet{Neureiter23} and~\citet{deNicola22b} we show that our triaxial models are able to recover the correct dynamical structure and mass composition of a realistic simulated ETG merger remnant with an accuracy at the 5\%-10\% level. 
\texttt{SMART} uses the density output from \texttt{SHAPE3D} to compute the corresponding stellar part of the gravitational potential by expansion into spherical harmonics. \texttt{SMART} launches stellar orbits, including near-Keplerian orbits in the center, from a 5-dimensional orbit starting space. It can use the entire information contained in the non-parametric LOSVDs.
\\
When modeling NGC\,5419 with \texttt{SMART} we focus on recovering its stellar orbit distribution, in particular its anisotropy distribution. We furthermore want to obtain updated measurements of the stellar mass-to-light ratio $\Upsilon$ and black hole mass $M_\mathrm{BH}$ of NGC\,5419. \\
For the modelling we assume that the central region of NGC\,5419 can be described by the potential of a single BH. We furthermore assume a constant mass-to-light ratio in our dynamical models and ignore possible initial mass function gradients. \\
We set up a parameter grid covering 10 stellar mass-to-light ratio values within $\Upsilon \in [4.0,8.0]$ with a corresponding grid size of $\Delta \Upsilon = 0.44$ and 10 tested black hole mass values within $M_\mathrm{BH} \in [0.5 ,2.0 ] \times 10^{10} M_\odot$ with $\Delta M_\mathrm{BH} = 0.17 \times 10^{10} M_\odot$.
We furthermore vary the viewing angles $\theta$, $\phi$ and $\psi$ in our models in correspondence with the SB density candidates, which are provided by \texttt{SHAPE3D}.

\texttt{SHAPE3D} is able to constrain the range of possible orientations of a triaxial galaxy based on photometric information alone by discarding deprojections where the 
radial profiles of the flattenings $p=\frac{b}{a}$ and $q=\frac{c}{a}$ (with $a$ being the semi major, $b$ the semi intermediate and $c$ the semi minor axis of a triaxial galaxy) are not smooth \citep{deNicola20,deNicola22_BCGs}. 
When deprojecting the observed SB of NGC\,5419 we apply an rms cutoff  $\mathrm{rms}=\sqrt{\langle \left(\ln ( I_{\text{obs}}/I_{\text{fit}})\right)^2 \rangle} \leq 0.013$ for the maximum discrepancy between the observed and modelled surface brightness. We furthermore discard all deprojections, where $p$ and $q$ deviate by more than 0.1 from their expected values due to radial changes in the order of the principal axes. With this, \texttt{SHAPE3D} provides 25 candidate orientations and respective luminosity profiles, which we probe with the dynamical models.

We assume a DM profile similar to a spherical NFW halo \citep{Navarro96}, yet with an inner density slope $\gamma=0$. We vary the density normalisation $\rho_0 \in [10^{7.6},10^{8.1}] M_\odot/\mathrm{kpc}^3$ at $r=10$kpc ($\Delta \mathrm{log_{10}}(\rho_0 [M_\odot/\mathrm{kpc}^3] = 0.08$) and use a fixed scale radius $r_s=40$ kpc. \\ 
Following triaxial symmetry we can separate the kinematic input data of NGC\,5419 along the galaxy's apparent minor axis (cf. Figure~\ref{fig:velmap}) in order to model two equivalent data-sets without loss of information. This provides us with an estimate of the statistical error of our modeling results (cf. \citealt{Neureiter23}).

\section{Results} \label{sec:Results}

\subsection{Kinematic recovery} \label{sec: Kinematic recovery}

\begin{figure*}
    \centering
    \includegraphics[width=1.0\textwidth]{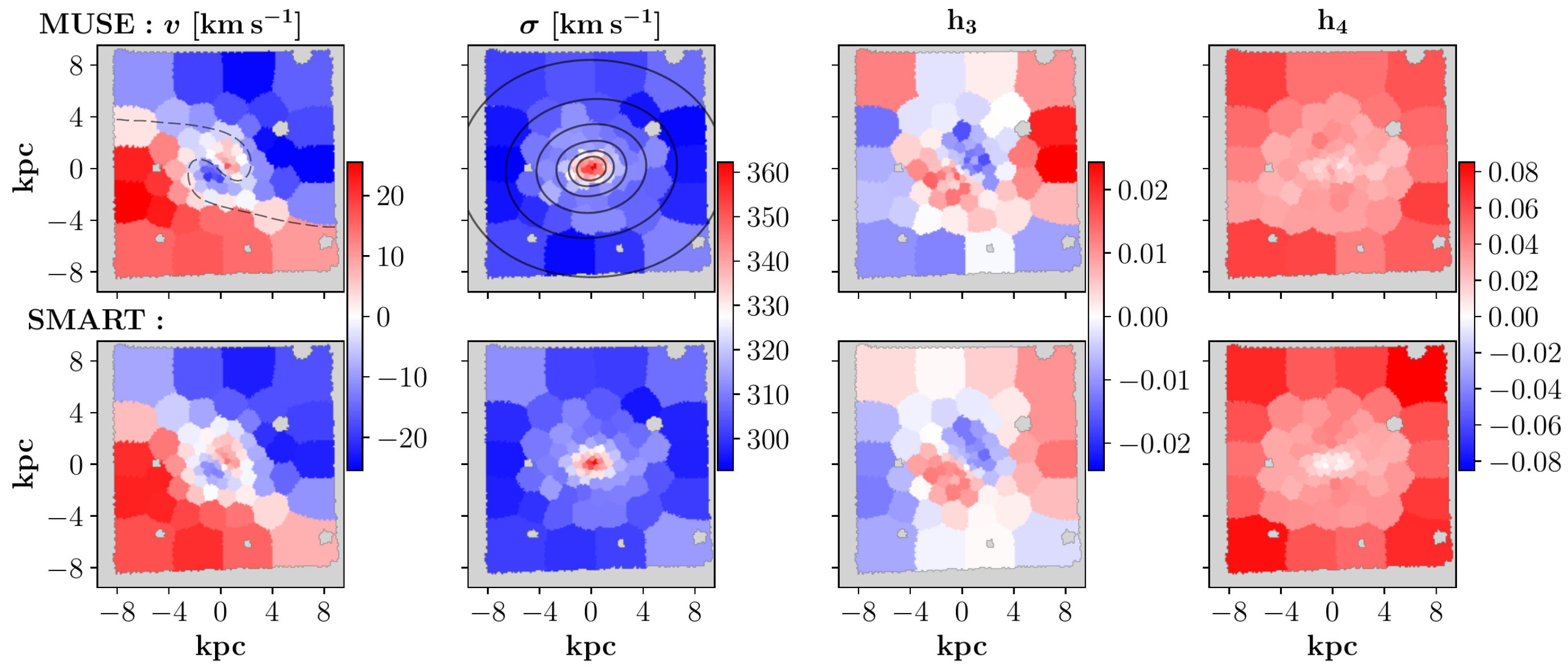}
    \caption{Velocity $v$, velocity dispersion $\sigma$ and Gauss-Hermite moments $h_3,h_4$ map of the MUSE data (top row) and accurately matching modeling fit by \texttt{SMART} (bottom row). \textcolor{black}{The observed velocity panel (top left) shows a prominent KDC.} \textcolor{black}{The counter-rotating core is misaligned by $\sim 120^\circ$ in comparison to the outermost velocity structure.} \textcolor{black}{The inner velocity pattern is not abruptly separated from the outer one but the kinematic axis rotates continously (see also Fig.~\ref{fig:SB}). This connection makes the overall velocity pattern resemble a 'yin-yang' symbol (illustratively sketched in the top left panel with the dashed curve). For comparison, NGC5419's isophotes are shown in black in the model velocity dispersion map (top second panel).} The MUSE kinematics are aligned along the position angle $\mathrm{PA}=78^\circ$ in agreement with the photometric major axis, i.e. North is at the top \textcolor{black}{and East points to the left} (cf.~\citealt{Mazzalay16,Mehrgan23}).}
    \label{fig:velmap}
\end{figure*}

\begin{figure}
    \centering
    \includegraphics[width=0.45\textwidth]{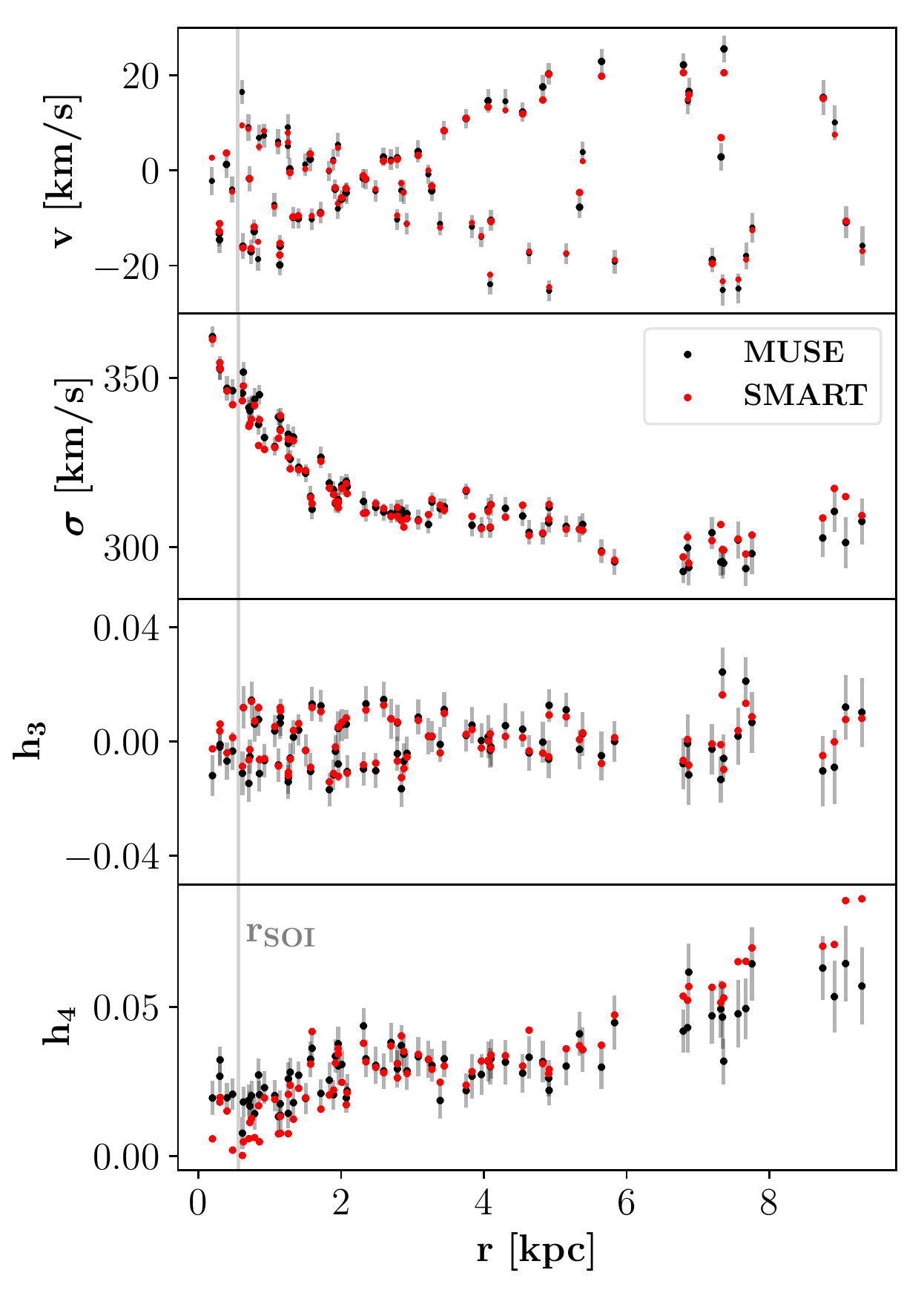}
    \caption{\textcolor{black}{Comparison of Gauss-Hermite parameters of the observed MUSE data (black points with error bars) and best-fit model (red points). This plot shows all data and model points as a function of radius computed for all Voronoi bins. 
    \textcolor{black}{The sphere-of-influence $r_\mathrm{SOI} (=0.57 \mathrm{kpc}$; defined as $M_*(r<r_\mathrm{SOI})=M_\mathrm{BH}$; see also Section~\ref{sec: Mass recovery}), is marked as grey vertical line.} While our triaxial model fits the full non-parametric LOSVDs, we here illustrate the quality of the fit in terms of Gauss-Hermite parameters that were determined a posteriori. The agreement between model and data is very good. }}
    \label{fig:v_sigma_h3_h4_radius_moments}
\end{figure}

Figure~\ref{fig:velmap} shows the MUSE kinematic maps in the top row and the best-fit model by \texttt{SMART} in the bottom row. The kinematic structure of NGC\,5419 can be well explained by the triaxial models (averaged over both sides we get $\chi^2/N_\mathrm{data}=0.45$). As already visible in the SALT stellar kinematics \citep{Mazzalay16} but much more clearly revealed by the MUSE observations \citep{Mehrgan23}, the velocity map of NGC\,5419 points to a kinematically distinct core (KDC).
We will come back to the KDC in Section~\ref{sec: Origin of KDC}. \\
\textcolor{black}{Figure~\ref{fig:v_sigma_h3_h4_radius_moments} shows the quality of the triaxial dynamical modeling fit illustrated by a comparison between the Gauss-Hermite parameterizations of the observed MUSE data (black data-points) and modelled fit (red data-points).} 
\textcolor{black}{We note that all the main features of NGC\,5419, in particular its specific velocity pattern including the KDC, can be explained in our equilibrium triaxial model. This suggests that the galaxy is observed at an evolutionary state, which is relaxed enough such that the dynamical modelling results are robust.}

\subsection{Shape recovery} \label{sec: Shape recovery}

\begin{figure}
    \centering
    \includegraphics[width=0.5\textwidth]{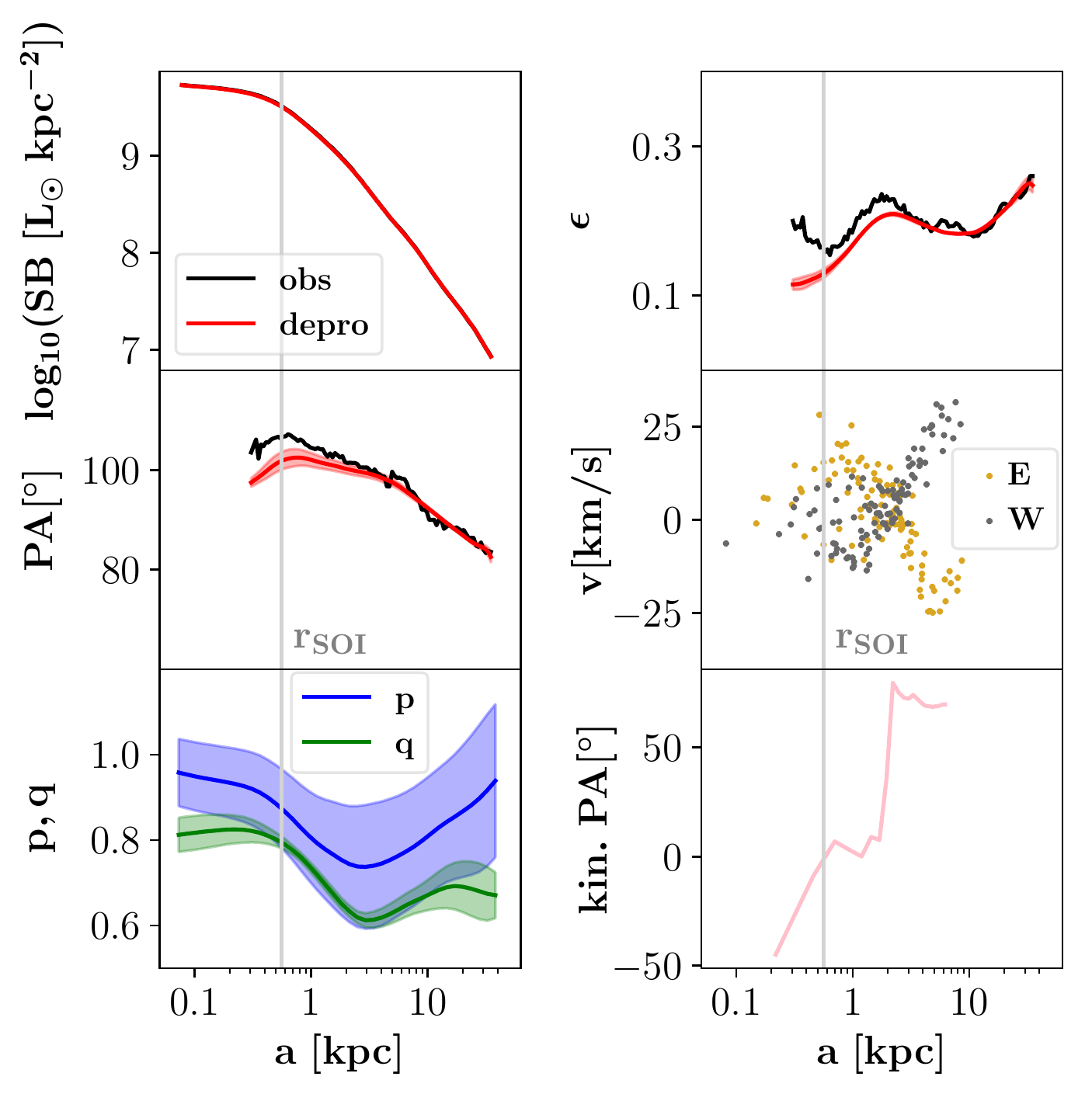}
    \caption{\textcolor{black}{Surface brightness $\mathrm{SB}$, ellipticity $\epsilon$, photometric position angle $\mathrm{PA}$, velocity $v$ along the galaxy's photometric major axis, axis ratios $p, q$ profiles and kinematic $\mathrm{PA}$ as a function of the elliptical isophotes' semi-major axis length $a$. The red solid line shows the mean and the shaded region shows the stardard deviation of the best-fit deprojections of the modelled two sides of the galaxy. The deprojection follows the observed profile (black line) within the relevant region of $r>r_\mathrm{SOI}$ (\textcolor{black}{marked as grey vertical line}). The bumps in the ellipticity profile match with the rotation sign changes in the velocity signal of the MUSE data along the galaxy's major axis (second right panel). The $p$ and $q$ profiles reveal a nearly oblate shape of the center and the outskirts of NGC\,5419. The variation in the kinematic $\mathrm{PA}$ corresponds to the observed KDC.}}
    \label{fig:SB}
\end{figure}

Figure~\ref{fig:SB} shows profiles of the observed surface brightness ($\mathrm{SB}$), ellipticity ($\epsilon$) and position angle ($\mathrm{PA}$) (colored black) as well as the best-fit deprojection (colored red).
The red shaded line is bounded by the two best-fit deprojections from modeling the two sides of the kinematic observational data (cf. Section~\ref{sec:Triaxial Dynamical Modeling}) and we also show their mean (red solid line).
The data suggest that the line-of-sight is between the galaxy's long and short axes with best-fit viewing angles\footnote{For a detailed explanation of the viewing angles, see \citet{Neureiter21}.} of $\theta=(65\pm10)^\circ, \phi=(30\pm10)^\circ, \psi=(100\pm10)^\circ$.
At this orientation both the SB profile and the kinematics of the triaxial model match very well with the observations at all radii. 
\textcolor{black}{The observed position angle shows a twist of $\sim20^\circ$, such that the galaxy cannot be axisymmetric. The relatively large $\mathrm{PA}$ twist combined with the observed ellipticity, which shows bumps of the order of $\Delta \epsilon \sim 0.05$, are consistent with the findings described in the following Sections~\ref{sec: Anisotropy recovery},~\ref{sec: Origin of KDC} and~\ref{sec: A core in formation?} pointing to a late but presumably not \textit{fully} completed merging phase of the galaxy.}
The deprojected $\epsilon$- and $\mathrm{PA}$-profiles match the observed ones particularly well outside the sphere-of-influence $r_\mathrm{SOI}$ (grey vertical line, c.f. Section~\ref{sec: Mass recovery}). \\
The radii, where the ellipticity profile features bumps, appear to match with the radii at which the rotation velocity along the galaxy's photometric major axis (aligned along $\mathrm{PA=78^\circ}$, see Figure~\ref{fig:velmap}) changes direction (see lower right panel in Figure~\ref{fig:SB} with the Eastern side of the kinematic data colored brown and the Western side colored grey). These radial changes in the rotation direction can also be seen in the velocity map in Figure~\ref{fig:velmap} and their origin will be explained in Section~\ref{sec: Origin of KDC}. \\
The bottom left panel in Figure~\ref{fig:SB} shows the $p$- and $q$-profiles of the best-fit deprojections of NGC\,5419, where one can see that this galaxy becomes oblate both in its center and its outskirts while it is triaxial at intermediate radii. \\
\textcolor{black}{The bottom right panel quantifies the variation of NGC\,5419's kinematic position angle (determined with the Kinemetry Code of \citealt{Krajnovic06}). The total amplitude of the variation is $\sim120^\circ$, in accordance with the visual impression that the centre of the galaxy 'almost' counter rotates with respect to the outer parts (Fig.~\ref{fig:velmap}). \\
Altoghether, the i) photometric twist, ii) variation in $\epsilon$ and iii) the change of the kinematic $\mathrm{PA}$ are typical for a triaxial galaxy.}

\subsection{Mass recovery} \label{sec: Mass recovery}
\begin{figure}
    \centering
    \includegraphics[width=0.5\textwidth]{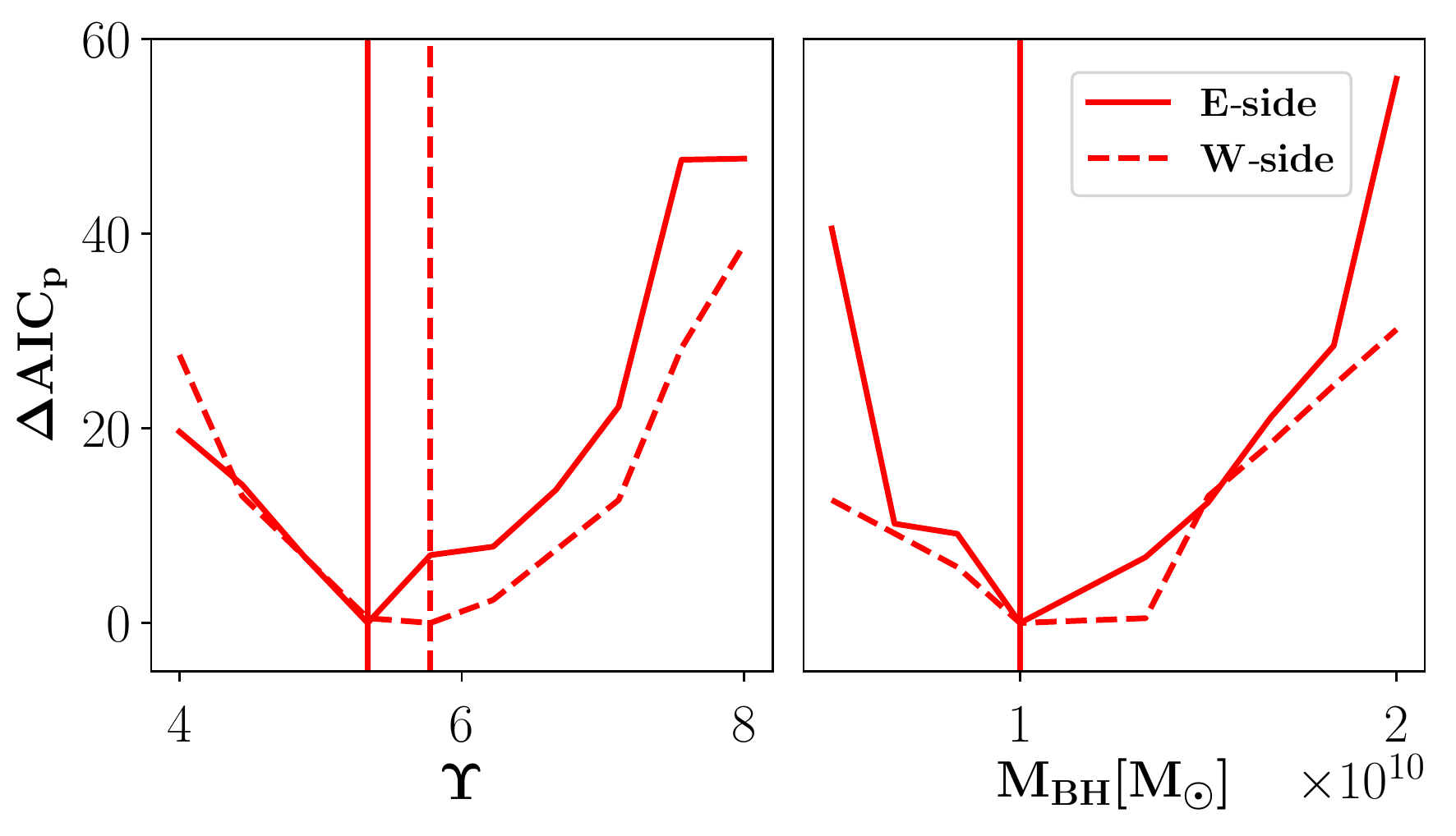}
    \caption{Minimum-subtracted $\mathrm{AIC_p}=\chi^2+2m_\mathrm{eff}$ curves for $\Upsilon$ (left panel) and $M_\mathrm{BH}$ (right panel) when modeling the Eastern (solid line) and Western side (dashed line) of the MUSE data with \texttt{SMART}. Instead of minimizing only $\chi^2$, our advanced dynamical modeling machinery takes the degrees of freedom $m_\mathrm{eff}$ of the individual models into account to avoid any bias when determining the best-fit parameters. The vertical lines show the best-fit mass parameters ($\Upsilon = 5.56 \pm 0.22 $ and  $M_{BH}= (1.0 \pm 0.08)\times 10^{10} M_\odot$).}
    \label{fig:mass}
\end{figure}

Figure~\ref{fig:mass} shows the minimum-subtracted $\mathrm{AIC_p}$ curves (cf. Section~\ref{sec:Triaxial Dynamical Modeling}) 
plotted against the stellar mass-to-light ratio $\Upsilon$ (left panel) and black hole mass $M_\mathrm{BH}$ (right panel). The solid line corresponds to the models covering the Eastern side and the dashed line corresponds to the models covering the Western side of the kinematic MUSE data. The vertical lines mark the mass parameters of the best-fit model, which is determined as the one with the minimum of all $\mathrm{AIC_p}$-values. \\
As Figure~\ref{fig:mass} shows, we obtain well-determined best-fit mass parameters.
Averaged over the two sides we measure $\Upsilon = 5.56 \pm 0.22 $ and $M_\mathrm{BH}= (1.0 \pm 0.08)\times 10^{10} M_\odot$. \\
With this, we determine a sphere of influence\footnote{We use the definition of the sphere of influence as the radius within which the total stellar mass equals the black hole mass.} of $r_\mathrm{SOI}=(0.57 \pm 0.02)$ kpc, i.e. $(2.09 \pm 0.07)$ arcsec, for NGC\,5419.
We find that our best-fit BH mass well fits into the $M_\mathrm{BH}$-$r_b$-correlation found by~\citet{Rusli13} based on 20 analysed core galaxies with dynamical $M_\mathrm{BH}$ measurements, confirming our results (we here use the core radius measurement $r_b=1.58$ arcsec from~\citealt{Mazzalay16}\textcolor{black}{; also the similar core radius measurement of $r_b=1.43$  arcsec from \citet{Dullo14} aligns with the $M_\mathrm{BH}$-$r_b$-correlation by~\citet{Rusli13} within its uncertainties).} 
The DM density normalization is recovered as $\mathrm{log_{10}}(\rho_0 [M_\odot/\mathrm{kpc}^3]) = 7.93 \pm 0.08$. \\
With a PSF-radius of $\sim 0.78$ arcsec for the MUSE data alone the sphere of influence of $\sim 2.09$ arcsec is marginally resolved. The constraints on $M_\mathrm{BH}$ are significant (the difference between the two modelled sides is very small), probably due to the high SNR ($\gtrsim 200$ per spectral bin in the optical region in each of about 220 Voronoi bins distributed over the whole FOV) in the MUSE data, which allows to extract LOSVDs with a high level of detail. \\
To test the robustness of our measurement, we performed an additional modeling analysis including the central AO assisted SINFONI data. While the focus of this cross-check is the recovery of NGC\,5419's central orbit distribution, which is of particular interest for our current study (see Sections~\ref{sec: Anisotropy recovery} and~\ref{sec: A core in formation?}), we also try to again recover $\Upsilon$ and $M_\mathrm{BH}$ using this combined data-set. \\
Assuming that the SINFONI FOV coverage is too small to significantly affect the DM halo recovery, we remodelled a 2D grid with mass variations of the black hole and stellar mass-to-light ratio and set the gNFW parameters to the best-fit DM halo parameters from modelling the MUSE kinematics only.
With this simplified analysis, \texttt{SMART} finds a best-fit stellar mass-to-light ratio of $\Upsilon=6.1 \pm 0.1$ and black hole mass of $M_\mathrm{BH}=(7.2\pm 0.5)\times 10^9 M_\odot$ with an average kinematic deviation of only $\chi^2/N\mathrm{data}=0.49$. With this, the so found BH mass deviates by $28$\% and the stellar mass-to-light ratio slightly deviates by $10$\% in comparison to the modeling results based on the MUSE data alone. More crucial, however, is the fact that the anisotropy profile stays unaffected by the used data-set and corresponding best-fit mass parameters (see Section~\ref{sec: Anisotropy recovery}). \\

\subsection{Anisotropy recovery} \label{sec: Anisotropy recovery}
\begin{figure}
    \centering
    \includegraphics[width=0.4\textwidth]{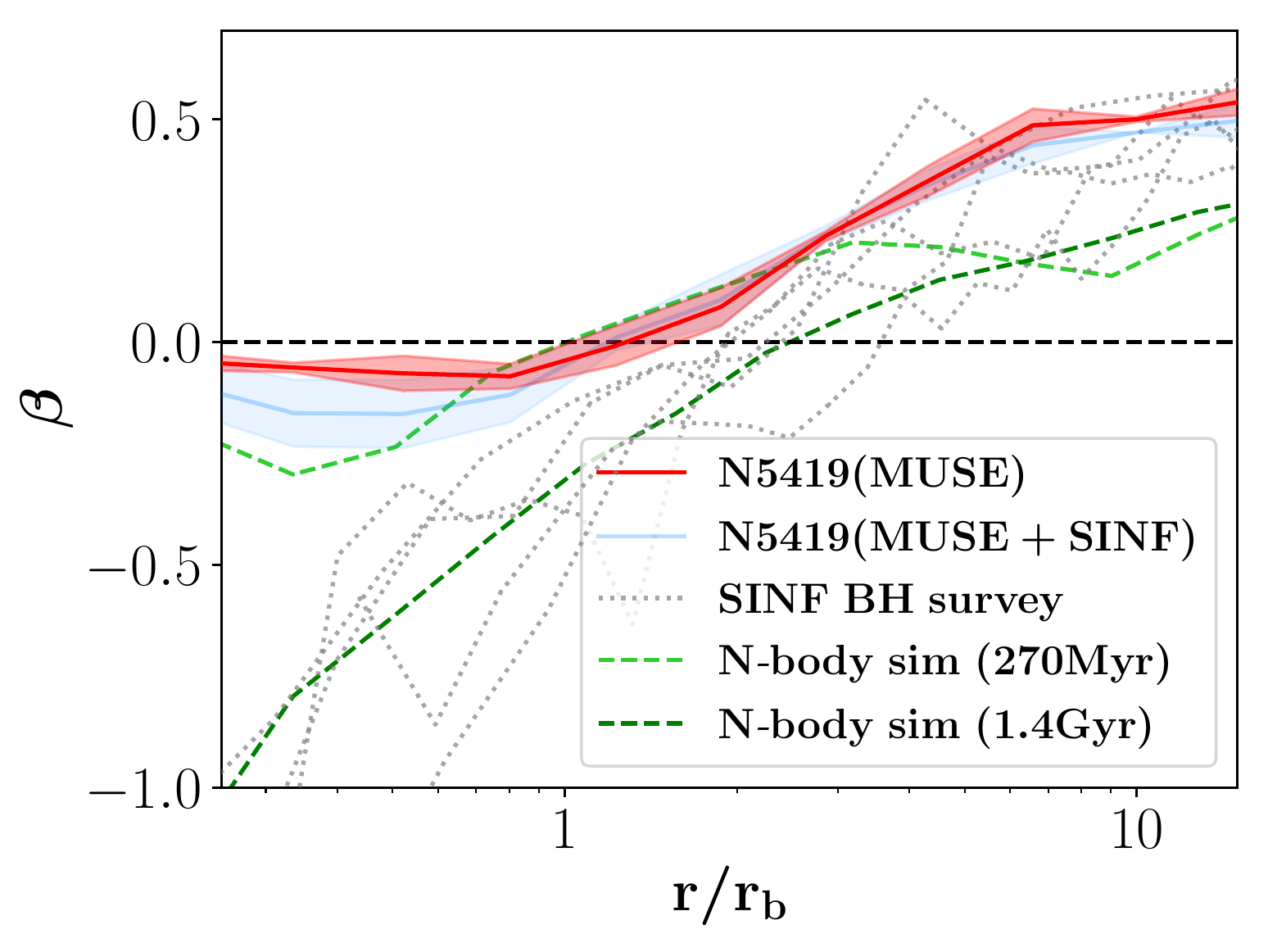}
    \caption{Anisotropy profile as a function of the radius scaled by $r_b$. 
    The red shaded region is bounded by the $\beta$-profiles when modeling the two sides of the MUSE data and we also show the mean of the two models in between (red solid line). In contrast to the dark green and grey dashed lines, which correspond to the $\beta$-profiles of a numerical gas-free major merger simulation with cuspy progenitor galaxies (observed $\sim 1.4$ Gyr after the merger has happened) and a sample of SINFONI-observed core ETGs, NGC\,5419 shows an isotropic central orbit distribution. This behavior is data-independent and does not change when modeling the MUSE kinematics combined with central high-resolution SINFONI data (blue line). The light green dashed line corresponds to the ETG merger simulation observed at an earlier stage ($\sim 270$ Myr after the merger) which will be explained and discussed in Section~\ref{sec: A core in formation?}.}
    \label{fig:anisotropy}
\end{figure}

Figure~\ref{fig:anisotropy} shows the mean anisotropy profile of the best-fit models of the two sides when modeling the MUSE data with \texttt{SMART}, plotted against the radius scaled by the core radius $r_b=1.58$ arcsec \citep{Mazzalay16}. The red shaded region is bounded by the two respective best-fit models.
The anisotropy parameter $\beta=1-\frac{\sigma_{\theta}^{2}+\sigma_{\phi}^{2}}{2 \sigma_{r}^{2}}$ (e.g. \citealt{Binney08}) consists of the velocity dispersions along the radial ($\sigma_r$) and tangential directions ($\sigma_\theta, \sigma_\phi$) and describes whether a galaxy's orbit distribution is radially anisotropic for $\beta>0$ or tangentially anisotropic for $\beta<0$. 
Our triaxial models reveal an isotropic, i.e. $\beta \sim 0$, central orbit distribution for NGC\,5419. Such an isotropic central $\beta$-profile is highly unusual for cored ETGs and we will discuss its implication further in Section~\ref{sec: A core in formation?}.\\
The central isotropy remains when we model the combined data set MUSE+SINFONI (even though the best-fit models for this run are weakly tangential in the very center; cf. blue line in Figure~\ref{fig:anisotropy}). 
As another robustness check we modified the smoothing parameter of the central deprojection. As visible in Figure~\ref{fig:SB}, our best-fit deprojections used for the dynamical models show minor deviations in the $\epsilon$- and $\mathrm{PA}$-profiles within $r_\mathrm{SOI}$. By reducing the smoothing in \texttt{SHAPE3D} (for a detailed description see~\citealt{deNicola20}), we obtained additional deprojections that better fit the central changes in $\epsilon$ and $\mathrm{PA}$. As expected, however, remodelling the galaxy with these deprojections (in our best-fit MUSE mass distribution) had no impact on the central anisotropy structure.

\subsection{Origin of the kinematically \textcolor{black}{distinct} core (KDC)} \label{sec: Origin of KDC}

\begin{figure*}
    \centering
    \includegraphics[width=0.92\textwidth]{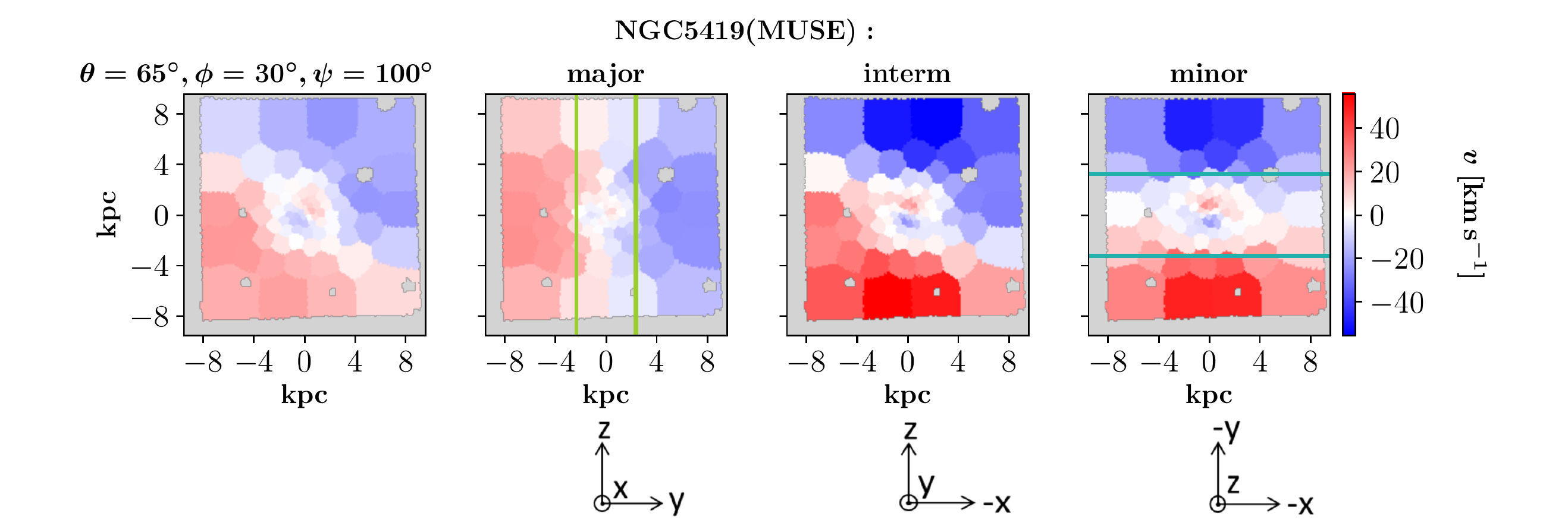}
    \caption{\textcolor{black}{Velocity maps of the best-fit model of NGC\,5419 projected along different line-of-sights: along the viewing angles assumed in the fit (left panel) as well as along the model's principal axes (second to fourth panel). The indicated coordinate systems demonstrate the orientation of the principal axes. The observed KDC and 'yin-yang velocity pattern' of NGC\,5419 appears to result from a projection effect caused by the overlay of z-tubes and x-tubes. The major (minor) axis projection singles out the rotation pattern of the z-tubes (x-tubes). 
    The stronger velocity signal is in the x-tubes while the velocity signal from the z-tubes is weaker but both show a velocity flip. Inside the flip-radius marked by green vertical lines, the z-tubes rotate in the opposite direction than in the outskirts and inside the blue horizontal lines, the x-tubes rotate in the opposite direction than outside. }}
    \label{fig:KDC}
\end{figure*}

\begin{figure*}
    \centering
    \includegraphics[width=0.6\textwidth]{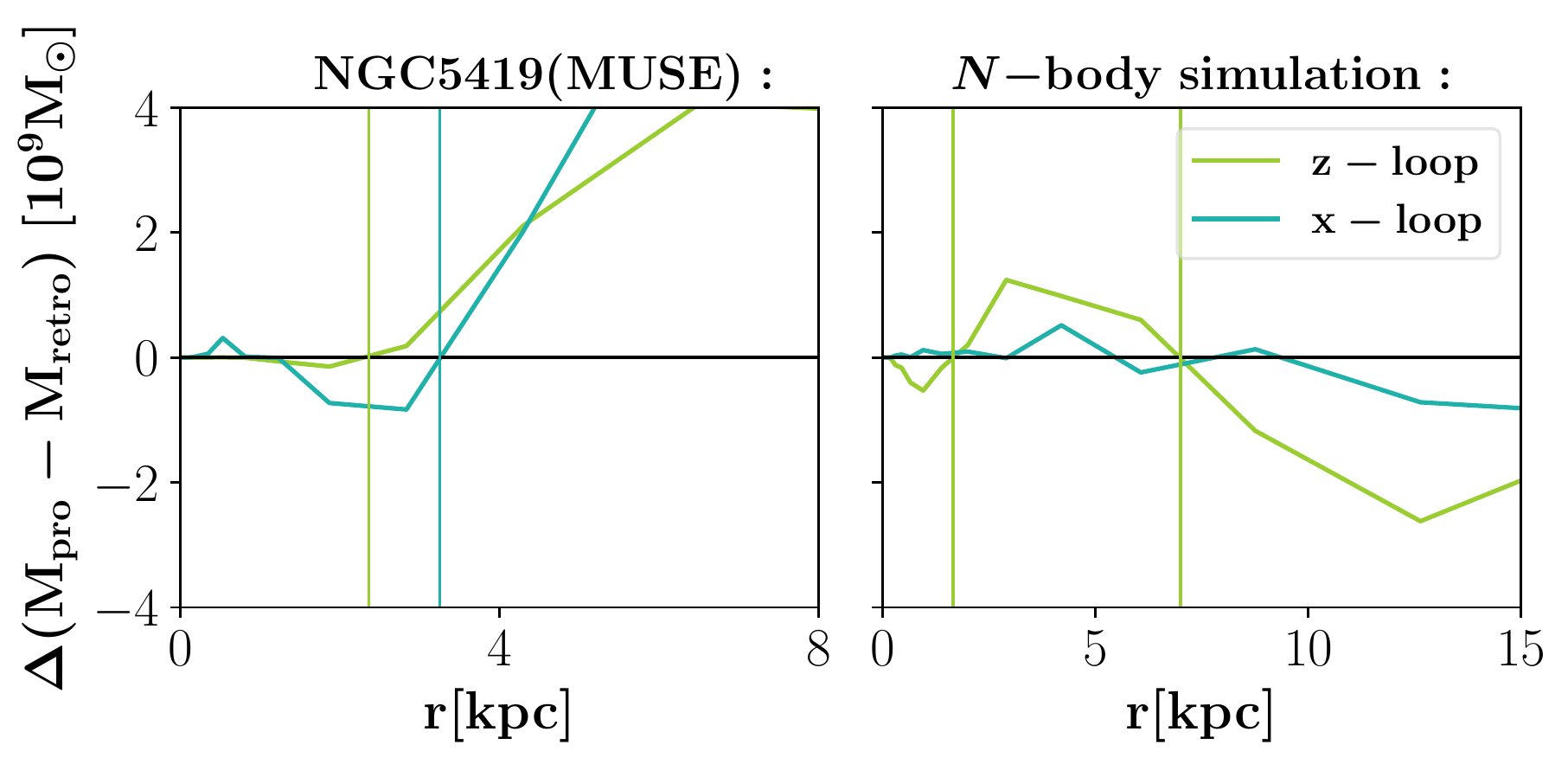}
    \caption{\textcolor{black}{Left: Mass excess of pro- and retrograde z- (green) and x-tube orbits (blue) in NGC\,5419. The radial regions of KDCs can be determined as "flip-radii" (vertical lines) at which the net rotation direction of a certain orbit family (z- and/or x-tubes) switches sign. The net rotation results from the prograde orbits carrying more stars than the retrograde ones or vice versa. Right: same for a merger simulation. While the $\Delta(M_\mathrm{pro}-M_\mathrm{retro})$-profile of the numerical merger simulation shows two flip-radii resulting in an inner and outer KDC, NGC\,5419 shows a single flip-radius. Nevertheless, the KDC in the simulation is -- similar to NGC\,5419 -- caused by the excess of tube orbits rotating around a preferred direction and the $N$-body simulation's outer flip-radius is qualitatively similar to the one of NGC\,5419. This hints to a similar formation process.  }}
    \label{fig:shellmass}
\end{figure*}

 As already shown in Figure~\ref{fig:velmap} and briefly described in Section~\ref{sec: Kinematic recovery}, NGC\,5419 shows a KDC. In general, KDCs together with isophote twists are frequently observed in massive ellipticals with depleted stellar cores like NGC\,5419 (e.g.~\citealt{Efstathiou82,Bender88_b,Franx88,Emsellem04,Hau06,McDermid06,Krajnovic11,Ene18}). 
 \textcolor{black}{The early standard model for KDCs in ETGs was early merging with dissipative formation of the KDC (see e.g. \citealt{Bender96,Davies01}). This is a still valid formation scenario for some ellipticals, like NGC\,5322 \citep{Bender88_b,Dullo18} or IC\,1459 \citep{Franx88,Cappellari02}, which show a fast rotating core with stars that counter-rotate in a disk on orbits that are close to circular and slow outer parts rotating in the opposite direction.  \\
Several observations suggest that the stellar populations of KDCs in some slowly rotating ETGs show little or no difference to the stellar populations of the surrounding host galaxies (\citealt{Davies01,McDermid06,Nedelchev19,Kuntschner10}).
Recent high-resolution gas-free numerical merger simulations hosting SMBHs (\citealt{Rantala19,Frigo21}) provide another possible explanation for the origin of kinematically distinct velocity structures in such ETGs. }
These simulations consist of collisionless dark matter halos, collisionless stellar components and SMBHs and they accurately compute the collisional interactions of the stars with the SMBHs with a regularized integration scheme (\citealt{Rantala17,Rantala18}).
In these simulations, the formation of counter-rotating cores is explained by 
the infall of the SMBHs with the bound stellar nuclei, which experience angular momentum reversals of their orbits after pericenter passages. The gravitational torque effects resulting in the reversals can be caused by deflections between the bulge and the halo due to the merger process \citep{Barnes16} or "dynamical self-friction", in which tidally expelled material exerts a force on the merger subhalo \citep{Bosch18}.
The specific velocity pattern of NGC\,5419 visible in Figure~\ref{fig:velmap}, however, not only shows a counter-rotating core but, more generally, a kinematically \textcolor{black}{misaligned} core in the sense that the inner velocity structure shows a slight misalignment in comparison to the direction of the outermost velocity structure. Moreover, the 2D velocity pattern resembles a kind of 'yin-yang symbol' on the sky caused by an apparent connection between the velocity pattern of the KDC and its surrounding large-scale velocity field. \textcolor{black}{The connections between the red and blue inner and outer velocity structures, which cause the similarity to a yin-yang symbol, become even more apparent with the finer bins of \citet{Mehrgan23}.} \\
Figure~\ref{fig:KDC} shows rotation maps of the best-fit model of the MUSE kinematics (when modeling the whole kinematic data with the best-fit parameters from the models for the Eastern side) projected along different line of sights: along the best-fit orientation of the galaxy (left panel), as well as along its three principal axes (labelled major, interm and minor). Our projected velocity maps reveal that the origin of such complex and misaligned KDCs as in NGC\,5419 can be simply explained by an overlay of orbits rotating around two different axes in an equilibrium triaxial model. These orbits are z-tubes, e.g. visible in the major axis projection, and x-tubes, e.g. visible in the minor axis projection (cf. Figure~\ref{fig:KDC}, second and fourth panel). 
\textcolor{black}{As one can see in Figure~\ref{fig:KDC}, the z-tubes within $r<2.4$ kpc (marked as green vertical lines) and the x-tubes within $r<3.3$ kpc (marked as blue horizontal lines) rotate in the opposite direction as their respective counterparts at large radii.} 
\textcolor{black}{The left panel in Figure~\ref{fig:shellmass} demonstrates this behaviour in another way by showing the radial mass difference $\Delta(M_\mathrm{pro}-M_\mathrm{retro})$ between prograde and retrograde z- and x-tube orbits, which is computed within radial shells. 
The orbit classification in our triaxial dynamical models is thereby done by checking the sign conservation of the angular momentum components along the three principal axes during the surfaces of section (SOS) crossings (for a more detailed explanation see~\citealt{Neureiter21}).
Outside the flip-radius $r=3.3$ kpc of the x-tubes, the prograde component dominates 
 and inside this radius the retrograde component dominates} \textcolor{black}{(we here define prograde orbits as the ones with $L_x>0$ and retrograde orbits as the ones with $L_x<0$ according to the coordinate systems shown in Figure~\ref{fig:KDC}). }\textcolor{black}{This is in agreement with the prominent central counter-rotation signal of the x-tubes visible in the minor-axis projection of our model of NGC\,5419 (cf. Figure~\ref{fig:KDC}, fourth panel).
Outside of $r=2.4$ kpc, the prograde component of the z-tubes dominates clearly over the retrograde component and within $r<2.4$ kpc, the components are almost equal with a slight overabdundance of retrograde z-tubes. This corresponds to the weak central counter-rotation signal of the z-tubes visible in the major-axis projection of our model of NGC\,5419 (cf. Figure~\ref{fig:KDC}, first row, second panel).}\\
\textcolor{black}{In order to understand whether the properties of NGC\,5419's KDC can be explained by the black-hole spin-flip scenario outlined above, we now compare it to the numerical $N$-body simulation by \citet{Rantala19} for which the counter-rotating core was shown to result from the rotation flips. 
This ETG major merger simulation includes two equal-mass SMBHs with a total black hole mass of $1.7 \times 10^{10} M_\odot$ and the two progenitor galaxies are set up by using a spherically symmetric Dehnen density-potential \citep{Dehnen93} with an initial inner stellar density slope of $\rho \propto r^{-3/2}$. 
In a series of recent papers (\citealt{Neureiter21,deNicola22b,Neureiter23}) we have modelled this simulation in great detail in order to test our triaxial modelling code. This $N$-body simulation was designed to explain another core galaxy (NGC\,1600) which is only slightly different from NGC\,5419 in terms of its stellar and black-hole mass. 
As shown in \citet{Rantala19}, the velocity amplitude of the KDC signal is qualitatively similar to the one observed in NGC\,5419. Different to NGC\,5419, this simulation has undergone a second reversal of the SMBH angular momentum. 
As one can see in Figure~\ref{fig:shellmass}, right panel, the rotation flip in the simulation is mainly caused by an overpopulation of retrograde z-loops within $r<r_1=1.65$ kpc, an overpopulation of prograde z-loops within $r_1 < r < r_2=7.0$ kpc and an overpopulation of retrograde z-loops in the outskirts $r>r_2$. The $N$-body simulation's outer flip-radius $r_2=7.0$ kpc thereby is qualitatively similar to NGC\,5419's single flip-radius.  \\
With this, we find that in both cases, in NGC\,5419 as well as in the $N$-body simulation, the KDC can be explained by the mass excess of counter-/corotating tube orbits inside/outside the respective flip radii.
While the number of rotation flips and the type of affected tube orbits can be expected to depend on the merger's specific initial conditions, like the orbit geometry and impact parameter, the spatial scale and velocity amplitude of NGC\,5419's KDC is notably similar to what can be found in the gas-free numerical merger simulation. \\
Alltogether this strongly suggests that the KDC in NGC\,5419 formed as a result of orbital angular momentum flips of the BHs during a merging process.} \\ \\
\section{Discussion} \label{sec: Discussion}
\textcolor{black}{\subsection{Comparison to other KDC studies}
Other studies from the literature, which analysed KDCs with Schwarzschild models (\citealt{Cappellari02,vandenBosch08,Krajnovic15,denBrok21}) suggested that their examined KDCs are not actually \textit{decoupled} in the sense that counter-rotating orbits are not exclusively localized in the central region but present everywhere. This is in agreement with our analysis from Section~\ref{sec: Origin of KDC}, where we show that NGC\,5419's KDC results from an excess of tube orbits rotating in the contrary direction compared to the outskirts.
A smooth excess mass profile as shown in Figure~\ref{fig:shellmass} is less likely for dissipationally formed KDCs in massive ETGs, such as in the already mentioned examples of NGC\,5322 or IC\,1459 (see Section~\ref{sec: Origin of KDC}), where the counter-rotating stars on a central disk have a more prominent separation in phase space. 
In this respect, the slow inner KDCs of some of the most massive galaxies might also be different from the ones in less massive galaxies, which are faster and probably also have a dissipative origin \citep{McDermid06}. The stellar populations in KDCs of less massive galaxies are often distinct from the rest of the galaxy (\citealt{Hau99,McDermid06,Nedelchev19}). The term kinematically \textit{decoupled} core might be misleading when used for both cases. 
For most massive galaxies, the term kinematically \textit{decoupled} core does neither infer an abrupt depletion of co-rotating orbits inside the KDC nor a distinct stellar population in the KDC. Nevertheless, the orbital properties inside the KDC are in some sense 'distinct' from the rest of the galaxy. The specific properties of the KDC likely depend on the specific dynamics that take place in the centre of a merger remnant and are somewhat local and detached from the global properties of the remnant. 
A probably more intuitive description of a "KDC" in slow rotating ETGs, as it is described in this study, might be a kinematically \textit{distinct} or \textit{misaligned} core (where a counter-rotating core would be a particular form of kinematical misalignment observed under certain viewing angles). 
}
\subsection{A core in formation?} \label{sec: A core in formation?}

NGC\,5419's isotropic central orbit distribution (Figure~\ref{fig:anisotropy} and Section~\ref{sec: Anisotropy recovery}) differs significantly from the strong tangential anisotropy observed in other core galaxies with high-resolution SINFONI data (\citealt{Thomas14}; grey lines in Figure~\ref{fig:anisotropy}). The orbital structure of these other core galaxies and in particular their central tangential anisotropy, however, is remarkably similar to the orbital structure in ETG merger simulations. \textcolor{black}{The dark-green dashed line plotted in Figure~\ref{fig:anisotropy} shows the $\beta$-profile of the $N$-body simulation by \citet{Rantala18}, which is observed $\sim 1.4$ Gyr after the merger has happened (we use $r_b(1.4\mathrm{Gyr})=0.49$kpc; \citealt{Rantala19}).} \\
Why is the core of NGC\,5419 isotropic?
As already explained in the Introduction, the core formation happens in two phases: First, the formation of a shallow central stellar density profile happens very rapidly within a time scale of some ten million years around the first passage of the two galaxies. This is followed by the much slower BH slingshot process that changes the central orbital structure from being isotropic to becoming tangential on a timescale of some 100 million years up to a Gyr. 
According to this, the ETG merger simulation snapshot $1.4$Gyr after the merger represents a galaxy where both formation phases have completed and the core region is strongly tangential. The cored ETGs in the sample of \citet{Thomas14} are all consistent with being ETGs where both core phases have already taken place. \\
A simple explanation for the fact that NGC\,5419 already has a flattened density profile but is still isotropic in its center would be that the galaxy is just between the two phases: the rapid dynamical-friction driven formation of the shallow density core has already passed yet the scouring process is just to begin. 
In fact, we analysed the same merger simulation \citep{Rantala18} at an earlier stage, $\sim 270$ Myr after the merger has happened, which is about the time when the BHs form a hard binary (\citet{Frigo21}; light green dashed line in Figure~\ref{fig:anisotropy}; we use $r_b(270\mathrm{Myr})=r_b(1.4\mathrm{Gyr})$). As one can see, at this earlier evolutionary stage, the center of the ETG merger simulation is indeed still isotropic (as already described in~\citealt{Rantala18} and~\citealt{Frigo21}). The similarity between the orbital structure of NGC\,5419 and the early ETG merger simulation snapshot supports the idea that core scouring is just at the beginning in NGC\,5419.\\
In principle, repeated minor mergers can also lead to a weaker central tangential anisotropy as compared to a single major merger \citep{Rantala19}, which can stay almost isotropic over time. However, the KDC in NGC\,5419 is characteristic for the orbital reversals of the progenitor SMBHs that only happen in major mergers with massive SMBHs \citep{Rantala19}. 
Another scenario presented in \citet{Rantala19}, which results in an only mildly tangentially biased central orbit distribution and velocity structure similar to NGC\,5419, would be a major merger of two core galaxies.
Here, violent relaxation processes and the infall of SMBHs can weaken the tangential anisotropy of the progenitor galaxies in the remnant. While in such a scenario the central orbit distribution can stay almost isotropic even after the BHs have merged~\citep{Rantala19}, a \textit{fully evolved} core-core merger scenario does not provide a direct explanation for the observed double nucleus.\\
In conclusion, the most plausible scenario is that while the galaxy merger, that forms NGC5419, is almost completed, the core formation process is at an \textit{earlier} state. This approach is able to simultaneously explain all observed and measured features:
\begin{itemize}[noitemsep,topsep=0pt]
    \item The central isotropic orbit distribution
\end{itemize}
\begin{itemize}[noitemsep,topsep=0pt]
    \item The double nucleus visible in the HST image surrounded by a region of enhanced stellar velocity dispersion 
\end{itemize}
\begin{itemize}[noitemsep,topsep=0pt]
      \item The already evolved cored SB profile and

      \item The galaxy's KDC, which probably reflects orbital reversals of the progenitor SMBHs that only happen in major mergers with massive SMBHs.
\end{itemize}
Finally, we note that the observed wiggles in the galaxy's ellipticity profile and their correlation with the flips in the rotation direction of the stars may further indicate that the galaxy is in an earlier phase of the core-formation process.
\textcolor{black}{Nevertheless, the evolutionary state of NGC\,5419 appears to already be relaxed enough in order to gain robust dynamical equilibrium models. This is supported by the plausible mass results and anisotropy profiles, which are similar for both modeled data-sets (Eastern and Western side of the kinematics), as well as by the fact that all specific velocity features can be well fitted with an averaged deviation of only $\chi^2/N_\mathrm{data}=0.45$.}


\subsection{Mass recovery comparison to other studies}
By modeling SINFONI data and optical long-slit SALT spectra assuming axisymmetry \citet{Mazzalay16} determined a best-fit stellar mass-to-light ratio of $\Upsilon=5.37^{+1.86}_{-1.42}$ and black hole mass of $M_\mathrm{BH}=7.24^{+2.74}_{-1.91}\times 10^9 M_\odot$. Within the error bars, their determined mass parameters agree with our best-fit values determined when modeling the MUSE data alone as well as with our best-fit parameters when modeling the combined MUSE and SINFONI kinematics (cf. Section~\ref{sec: Mass recovery}).

\section{Conclusion} \label{sec: Conclusion}
We present new triaxial dynamical models based on new spectroscopic MUSE data of the massive elliptical core galaxy NGC\,5419 which hosts a double nucleus.
We focused on a precise recovery of the galaxy's orbital structure. Our best-fit model shows a surprisingly isotropic central orbit distribution, which is in contrast to other core ETGs which typically show a tangentially biased central orbit distribution. 
The observed double nucleus together with the isotropic central orbit distribution most likely suggest that NGC\,5419 has undergone only the first phase of core formation so far: the dynamical friction associated with the sinking SMBHs in a gas-free merger causes a flattening of the central density profile. Such a depleted stellar density core is definitely in place in NGC\,5419. However, the subsequent SMBH scouring process, which kicks out stars on radial orbits and slowly produces a tangentially biased orbit distribution has probably just begun. This supports the idea that the double nucleus at the center of the galaxy is in fact a SMBH binary \citep{Mazzalay16}. \\
We find a best-fit stellar mass-to-light ratio of $\Upsilon = 5.56 \pm 0.22 $ and best-fit black hole mass of $M_{BH}= (1.0 \pm 0.08)\times 10^{10} M_\odot$ for the MUSE data-set. 
We find a slightly smaller best-fit BH mass when adding high-resolution AO-assisted SINFONI data.
The recovery of the anisotropy profile, however, is unaffected by this and in particular the central orbit distribution stays isotropic independent of which data is fitted. 

NGC\,5419 has a prominent yin-yang like KDC and is a particularly interesting case to study the formation of kinematic misalignements.  
When projecting our best-fit triaxial model along the galaxy’s principal axes, we find rotation flips in the tube orbits around \textcolor{black}{$r\sim3$kpc}. These flips together with projection effects are sufficient to produce the observed complex rotation field. 
Recent ETG merger simulations have shown that during gas-free major mergers, SMBHs (in the mass range implied by our models for NGC\,5419) and the bound stellar nuclei can experience one or more reversals of their orbits \citep{Rantala19}. 
\textcolor{black}{Similar to what can be found in our models of NGC\,5419, the traces of these orbital reversals in the $N$-body simulations appear as excess mass of counter-/corotating tube orbits inside/outside the effective flip radii. The fact that the spatial scale and velocity amplitude of NGC\,5419's KDC is qualitatively similar to what can be found in the $N$-body simulation hints to a similar formation process of the KDC.}

\section*{Acknowledgements}
This research was supported by the Excellence Cluster ORIGINS which is funded by the Deutsche Forschungsgemeinschaft (DFG, German Research Foundation) under Germany's Excellence Strategy - EXC-2094-390783311. We used the computing facilities of the Computational Center for Particle and Astrophysics (C2PAP). Computations were performed on the HPC systems Raven and Cobra at the Max Planck Computing and Data Facility.

\bibliography{bib}

\end{document}